\documentclass[aps,amsmath,amssymb,superscriptaddress,prb,reprint,twocolumn]{revtex4-2}
\usepackage{graphicx}
\usepackage{float}
\usepackage{dcolumn}
\usepackage{multirow}
\usepackage{braket, bm}
\usepackage[utf8]{inputenc}
\usepackage[T1]{fontenc}
\usepackage{mathptmx}
\usepackage{tabularx}
\usepackage{}	

\begin{document}
\title{Topological Edge and Corner States in Biphenylene Network}
\author{Keiki Koizumi}
\affiliation{Department of Nanotechnology for Sustainable Energy, School
of Science and Technology, Kwansei Gakuin University, Gakuen-Uegahara 1, Sanda 669-1330, Japan}
\author{Huyen Thanh Phan}
\affiliation{Department of Nanotechnology for Sustainable Energy, School
of Science and Technology, Kwansei Gakuin University, Gakuen-Uegahara 1, Sanda 669-1330, Japan}
\author{Kento Nishigomi}
\affiliation{Department of Nanotechnology for Sustainable Energy, School
of Science and Technology, Kwansei Gakuin University, Gakuen-Uegahara 1, Sanda 669-1330, Japan}
\author{Katsunori Wakabayashi}
\affiliation{Department of Nanotechnology for Sustainable Energy, School
of Science and Technology, Kwansei Gakuin University, Gakuen-Uegahara 1, Sanda 669-1330, Japan}
\affiliation{National Institute for Materials Science (NIMS), Namiki 1-1, Tsukuba 305-0044, Japan}
\affiliation{Center for Spintronics Research Network (CSRN), Osaka
University, Toyonaka 560-8531, Japan}
\date{\today}

\begin{abstract}
The electronic states and topological properties of the biphenylene
 network (BPN) are analyzed using a tight-binding model based on the
$\pi$-electron network. It is shown that tuning the hopping parameters
 induces topological phase transitions, leading to the 
 emergence of edge states owing to the nontrivial topological Zak phase of
 the bulk BPN. Elementary band analysis clearly gives the number of edge
 states, which are associated with the location of Wannier centers.
In addition, we have presented the conditions for the emergence of
 corner states owing to the higher-order topological nature of BPN.
\end{abstract} 

\maketitle

\section{Introduction}
The concept of topology has brought about the paradigm shift in modern condensed
matter physics, opening up the research field of topological materials. 
These materials include topological insulators~\cite{Kane2005,Bernevig2006,Fu2007,Konig2007,RevModPhys.82.3045,RevModPhys.83.1057,Hsieh2008,Chen2009,Chang2013,Ando2013}, 
topological crystalline insulators~\cite{Fu2011,Dziawa2012,Ando2015},
topological semimetals~\cite{Wan2011, Burkov2011, Borisenko2014,  
Liang2016, Watanabe2016, Yang2017} and so on.
One of the essential properties of topologically nontrivial systems is
the bulk-edge correspondence, where robust edge and surface states
appear at the interfaces that separate two topologically distinct systems. 
These topologically protected states are insensitive to local perturbation
such edge roughness and disorder, which may serve for the application
to the ultra-low-power-consumption electronics and quantum computation. 
The idea of device design based on topology has also been recently applied
to photonics~\cite{Raghu2008,Wang2008,Khanikaev2013,tp2014review} and
other systems~\cite{tphonon2020review,tcircuitreview}.

In a system with both time reversal and crystal inversion
symmetries, the Zak phase can serve as a good topological invariant
for characterizing the bulk-edge correspondence~\cite{Zak1989}. Since
the Zak phase is associated with charge
polarization~\cite{King-Smith1993,Resta1994,Marzari2012,Zhou2015,Liu2017},
a nontrivial finite Zak phase predicts the presence of edge-localized states when an edge is
introduced to the system. In graphene, a one-atomic thickness two-dimensional (2D) carbon
sheet, the edges induce the electron localized states (edge states) at the Fermi
energy~\cite{Fujita1996,Nakada1996,Wakabayashi1999}. However, the edge states crucially
depend on the shape of edges and are absent for armchair edges. 
This puzzle can be resolved by considering the Zak phase of the bulk wave
function of graphene, which gives the momentum dependent Zak phase~\cite{Ryu2002,Delplace2011}.
The Zak phase is identically zero solely for the armchair edge, i.e., no
edge states, which is consistent with various numerical calculations~\cite{Wakabayashi2009Carbon,wakabayashi2010jpsj}.
The presence of edge states provides the spin polarization~\cite{Fujita1996,ywson2006prl} 
and a perfectly conducting channel~\cite{Wakabayashi2007PRL}.

Recently, biphenylene network (BPN), a newly synthesized 2D sp$^2$-carbon-based material, 
was successfully synthesized by Fan {\it et.al.}~\cite{Qitang2021}. 
BPN has a fascinating lattice structure, where the hexagonal carbon rings are
organized on a square lattice, 
resulting in a 2D tiling pattern that includes four-, six-, and eight-membered rings.
Previous studies have reported thermal conductivity, magnetic properties, and hydrogen storage properties 
based on first-principles calculations~\cite{Bafekry2022, Luo2021, Son2022, Ma2024}.
Also, Ref.~\cite{Son2022} reported Zak phases and topological grain
boundary states along BPN nanoribbons. However, the higher order
topological corner states are not considered yet. Here we also employ
the elementary band analysis which clearly gives the number of edge
states, which are associated with the location of Wannier centers.


In this paper, we theoretically study the electronic and topological 
properties of 2D BPN. 
Since BPN comprises only $sp^{2}$ carbon atoms similar to
graphene, the $\pi$-electrons govern the electronic states of BPN near
the Fermi energy. Thus, we employ the tight-biding model, which describes
$\pi$-electronic states of 2D BPN to analyze the topological properties.
We will demonstrate that successive topological phase transitions occur
by tuning the ratio between intra- and inter-cellular electron hoppings.
In Su-Schrieffer-Heeger (SSH) model~\cite{Su1980,Heeger1988},
similar topological phase transition can occur, inducing nontrivial
Zak phase, resulting in the edge states~\cite{Liu2017_2,Obana2019,Pershoguba2012,Delplace2011}. 
We also find that the topological properties of BPN are characterized by 
the Zak phase of the bulk wave function. 
Charge polarization at the surfaces, i.e., edge states, is induced if the Zak phase possesses $\pi$. 
By studying BPN nanoribbons, we attribute the appearance of edge states
to the Zak phase of $\pi$. 
Furthermore, we have demonstrated that the exact number of edge states can be determined
by considering the position of the Wannier orbital. 
Similar arguments can be used to show the existence of topological corner states. 

This paper is organized as follows. In Sec.~\ref{sec:elec}, we
investigate the electronic states of $\pi$-electrons of BPN using
a tight-binding model. Considering two types of hopping energies, we show
that the topological phase transition occurs at a specific ratio. In
Sec.~\ref{sec:topo}, we illustrated the existence of the edge states in
BPN guaranteed by nontrivial Zak phase and the central position of
Wannier function. In addition, we find multiple corner states in BPN nanoflake. 
Section~\ref{sec:conclusion} provides the summary of the paper.

\section{Electronic states of 2D BPN}\label{sec:elec}
\subsection{Tight-binding model}
\begin{figure}[h]
\centering
\includegraphics[width=0.95\linewidth]{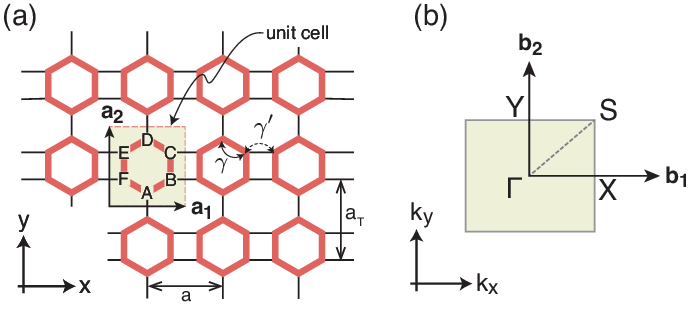}
\caption{
  (a) The lattice structure of BPN. The shaded region is the unit cell,
 with six nonequivalent sublattices of carbon atoms (A to F). $\bm{a}_1
 = (a, 0)$ and $\bm{a}_2 = (0, a_T)$ are two primitive vectors, where
 $a$ and $a_T=\frac{3a}{\sqrt{3}+1}$. The intra- (inter-) cellular
 hoppings denoted by red (black) lines are $\gamma$ ($\gamma^{\prime}$).  
  (b) The 1st BZ of BPN and their high-symmetric points, 
$\rm{\Gamma} = (0, 0)$, $\rm{X} = (\pi/a, 0)$, $\rm{Y} = (0,
 \pi/a_T)$, and $\rm{S} = (\pi/a, \pi/a_T)$.
  }
\label{fig:lattice}
\end{figure}
\begin{figure*}[hbt]
  \centering
   \includegraphics[width=0.95\textwidth]{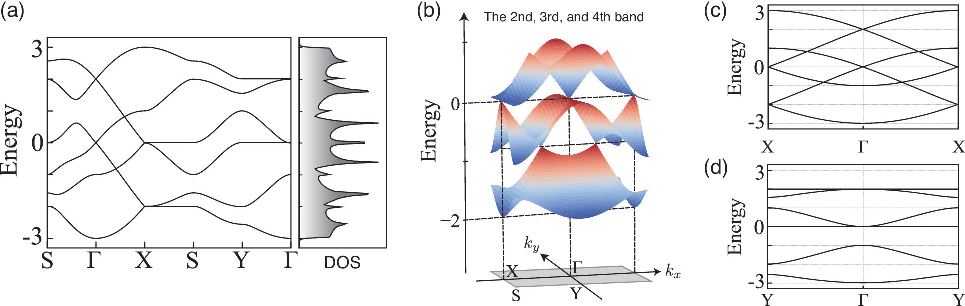}
  \caption{
    (a) The energy band structure and the corresponding DOS
        for $\gamma^{\prime}/\gamma = 1$. There are six subbands.
    (b) 3D plot of the energy band structure for the band-$2$, the band-$3$, and the band-$4$. 
    (c) The energy band structure along $k_{x}$ direction. 
        Tilted Dirac dispersions can be seen on
        $\Gamma$-$\rm{X}$ line between the band-$2$ and the band-$3$. 
    (d) The energy band structure along $k_{y}$ direction. 
        The band-$3$ is flat along the $\rm{\Gamma}$-$\rm{Y}$ line, and 
        the band-$4$ has a parabolic dispersion at $\rm{\Gamma}$ point. 
    }
  \label{fig:band_3d}
\end{figure*}
Figure~\ref{fig:lattice} (a) displays the schematic of the lattice structure
for 2D BPN, where the hexagonal carbon rings are
organized on a square lattice,
resulting in a 2D tiling pattern that includes four-, six-, and eight-membered rings. 
The shaded rectangle in Fig.~\ref{fig:lattice} (a) represents the unit cell. 
Since the BPN is composed of sp$^2$ carbon atoms, the electronic states
of 2D BPN near the Fermi energy are governed by $\pi$-electrons.
Now let us consider the nearest-neighbor tight-binding model for
$\pi$-electrons of 2D BPN. 
BPN contains six geometrically nonequivalent carbon atoms (A, B, C, D, E, F) 
in the unit cell as shown in Fig.~\ref{fig:lattice} (a). 
Here, $\bm{a}_{1} = (a, 0)$ and $\bm{a}_{2} = \left(0, a_T\right)$ 
are the primitive vectors. $a$ and $a_T=\frac{3a}{\sqrt{3}+1}$ are
lattice periodicities along $x$- and $y$-directions, respectively. 
The length of $a$ is $3.7554$\AA. 
We shall introduce two kinds of electron hoppings: 
$\gamma$ is the intra-cellular hopping, and $\gamma^{\prime}$ is the
inter-cellular hopping, respectively.
Figure~\ref{fig:lattice} (b) shows the first Brillouin
zone (BZ) with high-symmetric points, i.e., 
$\rm{\Gamma} = (0, 0)$, $\rm{X} = (\frac{\pi}{a}, 0)$,
$\rm{Y} = (0, \frac{\pi}{a_T})$, and
$\rm{S} = (\frac{\pi}{a}, \frac{\pi}{a_T})$.
The actual parameter set to reproduce the energy band structure for
$\pi$-electrons of 2D BPN is given in Appendix A.

The eigenvalue equation of 2D BPN is described as 
\begin{equation}
 \hat{H}(\bm{k}) \ket{u_{n, \bm{k}}} = E_{n, \bm{k}} \ket{u_{n, \bm{k}}}, 
\end{equation}
where $\hat{H}(\bm{k})$ is the Hamiltonian at the wave number $\bm{k}=(k_x, k_y)$, 
and $E_{n, \bm{k}}$ is the eigenvalue with the band index $n$ ($= 1, 2, \cdots, 6$).
The eigenvector $\ket{u_{n, \bm{k}}}$ is written by
\begin{equation}
 \ket{u_{n, \bm{k}}} = [c_{n, A}(\bm{k}), c_{n, B}(\bm{k}), c_{n, C}(\bm{k}), 
 c_{n, D}(\bm{k}), c_{n, E}(\bm{k}), c_{n, F}(\bm{k})]^T, 
\nonumber
\end{equation}
where $[\cdots]^T$ means the transpose of vector. 
$c_{n, \alpha}(\bm{k})$ is the amplitude of the wave function 
at the site $\alpha$ for the $n$-th band. 
The tight-binding Hamiltonian up to the
nearest-neighbor hopping is 
\begin{equation}
 \hat{H}(\bm{k}) = -\gamma 
  \begin{pmatrix}
   0 & 1 & 0 & \rho_{y} & 0 & 1\\
   1 & 0 & 1 & 0 & 0 & \rho^{*}_{x}\\
   0 & 1 & 0 & 1 & \rho^{*}_{x} & 0\\
   \rho^{*}_{y} & 0 & 1 & 0 & 1 & 0\\
   0 & 0 & \rho_{x}     & 1 & 0 & 1\\
   1 & \rho_{x} & 0     & 0 & 1 & 0\\
  \end{pmatrix}, 
\end{equation}
with
\begin{equation}
 \rho_{x} = \frac{\gamma^{\prime}}{\gamma}e^{i\bm{k} \cdot \bm{a}_1},\,
 \rho_{y} = \frac{\gamma^{\prime}}{\gamma}e^{i\bm{k} \cdot \bm{a}_2}.
\end{equation}

Figure~\ref{fig:band_3d} (a) shows the energy band structure along the path
connecting high-symmetric points of the 1st BZ, 
and the density of states (DOS) for $\gamma^{\prime}/\gamma = 1$.
Since BPN has a $\pi$-electron per atomic site in average,
the Fermi energy is $E=0$. 
There are six energy bands, and from the bottom to the top, and the
corresponding band index is $1$ to $6$.
BPN is metallic with flat bands at $E=0$. 
Also, owing to the crystal symmetry of BPN, the energy dispersion has
the following relation,
\begin{align}
E_n(k_x, k_y)& = -E_{6-(n-1)}\left(k_x\pm\frac{\pi}{a}, k_y\right). 
\end{align}

Figure~\ref{fig:band_3d} (b) is the 3D plot of the energy band structure 
in the 1st BZ. 
It should be noted that the band touchings occur 
only on the the line of $\Gamma$-$X$.
Figures~\ref{fig:band_3d} (c) and (d) show the energy band structure 
on $\Gamma$-$X$ and $\Gamma$-$Y$ lines, respectively. 
The 2nd and the 3rd subbands linearly cross at a point
on $\Gamma-X$ line, 
i.e., tilted Dirac cone. 
Similarly, the tilted Dirac cones also emerge between the $4$th and
the $5$th subbands.
However, the $3$rd and the $4$th subbands touch differently at $E=0$, i.e., 
isotropic energy dispersion.
At $\Gamma$ point, these bands linearly touch along the $k_{x}$ direction,
whereas the $3$rd subband becomes flat, 
and the $4$th subband has a parabolic dispersion in $k_{y}$ direction. 
Additionally, at $X$ point, the $3$rd subband is parabolic along $X$-$S$ line, while
the $4$th subband is almost flat along $X$-$S$ line.

\subsection{Effect of variable hopping energy}
\begin{figure*}[ht]
  \centering
  \includegraphics[width=0.9\textwidth]{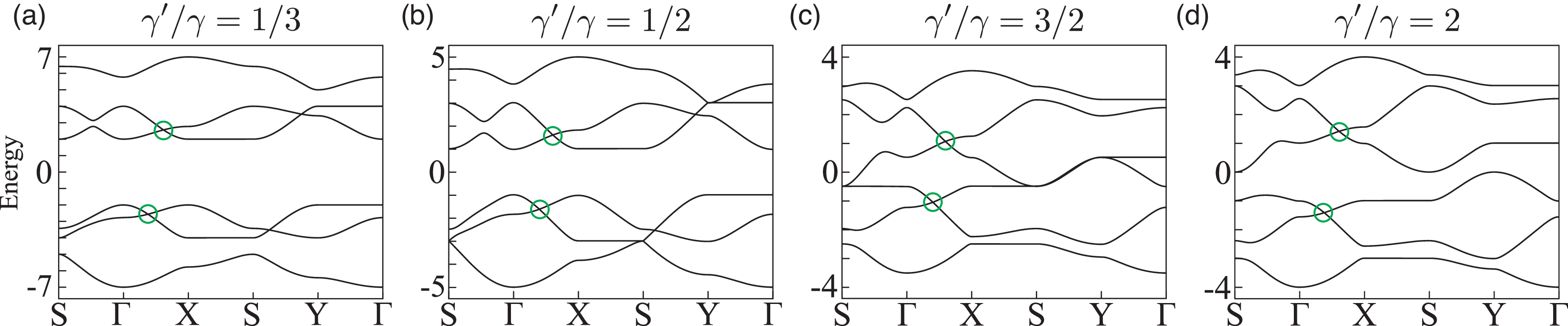}
  \caption{Energy band structure of 2D BPN for (a) $\gamma^{\prime}/\gamma = 1/3$,
    (b) $\gamma^{\prime}/\gamma = 1/2$, (c) $\gamma^{\prime}/\gamma = 1.5$, and
    (d) $\gamma^{\prime}/\gamma = 2$. The band degenerating points are marked with green circles.
    }
  \label{fig:band_var}    
  \end{figure*}

Figures~\ref{fig:band_var} (a)-(d) show the energy band structures of 2D
BPN with several different ratios of hopping energy
$\gamma^{\prime}/\gamma$. It can be seen that the topology of
the energy band structure crucially depends on the ratio of these two
hopping parameters. In BPN, the topological phase transition occurs at
$\gamma^{\prime}/\gamma = 1/2, 1$, and $1.5$, where the band gap closes
and reopens.  
The closing and reopening of the energy band gap cause a band inversion,
resulting in a topological phase transition, i.e., a change in the value
of the Zak phase. 

In the range of $1/2 \leq \gamma^{\prime}/\gamma \leq 1$, the 1st
subband touches the 2nd subband, similarly, the 5th subband touches the
6th subband. However, these band touchings are lifted for the other 
range and become isolated. 

Furthermore, the 3rd and the 4th subbands touch at the Fermi energy in the
range of $1 \leq \gamma^{\prime}/\gamma \leq 1.5$. Therefore, a complete
band gap opens when $\gamma^{\prime}/\gamma < 1$ and
$\gamma^{\prime}/\gamma > 1.5$. 
There are four tilted Dirac cones on the $\Gamma$-$X$ line; one is
between the $2$nd and the 3rd subbands, and another one is between the
$4$th and the $5$ subbands. These crossing points are marked by green circles in
Fig.~\ref{fig:band_var}. These four points are maintained in all cases.  

\section{Topological properties}\label{sec:topo}
\subsection{Zak Phase and Topological Edge States}\label{sec:Zak}
\begin{figure*}[t]
\centering
\includegraphics[width=0.95\textwidth]{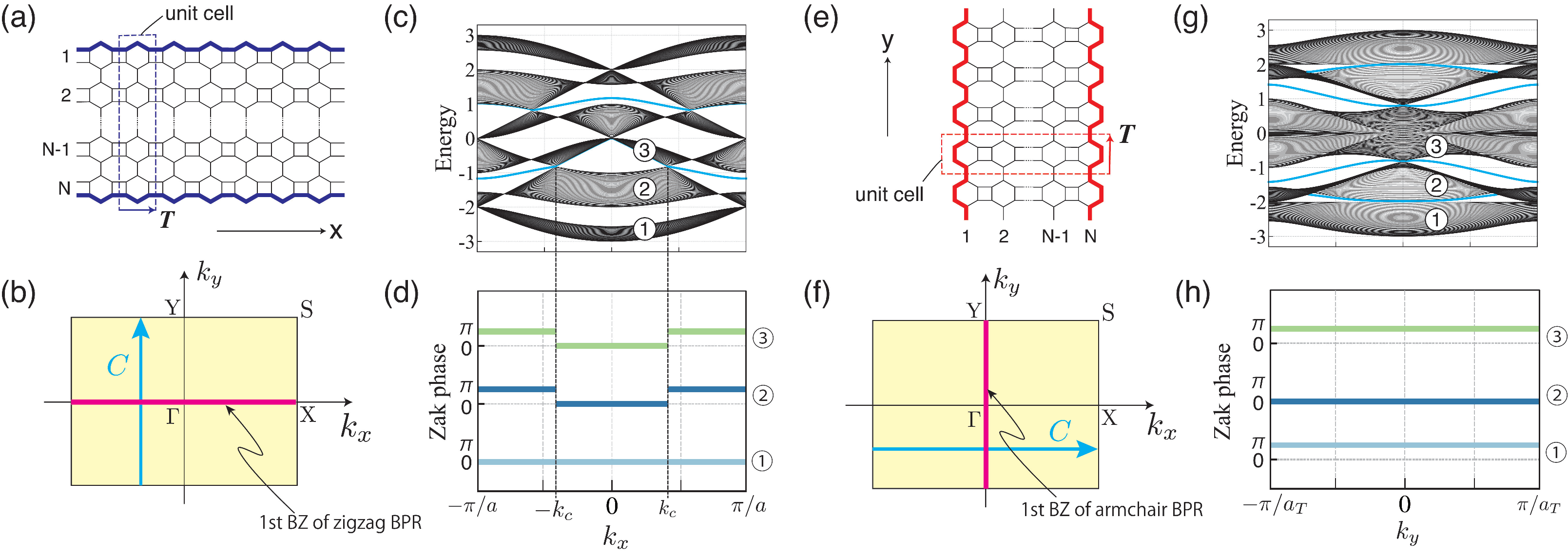}
 \caption{(a) Schematic lattice structure of BPN ribbon (BPR) with zigzag edges. 
$N$ is the width of the ribbon. The translational vector $\bm{T}$ for zigzag BPR is
 defined as $\bm{T} = T(1, 0)$. Here, 
$\bm{T} = m\bm{a}_1 + n\bm{a}_2 = T(m, n)$.
(b) The 1st BZ of zigzag BPR (thick magenta line). The shaded yellow area is
 the 1st BZ of 2D BPN. 
In zigzag BPR, its 1st BZ is projected
 onto $k_x$-axis and given as $-\pi/a\le k_x\le \pi/a$. 
The integration path $C$ for the calculation of the Zak phase is taken along
 $k_y$.
(c) Energy band structure for zigzag BPR for $N=50$. 
Cyan lines indicate the modes of TES that emerge in the energy gaps
 where the total Zak phase is $\pi$. 
(d) The corresponding Zak phase of zigzag BPR for the lowest three subbands,
where Zak phase is finite in the $2$nd and the $3$rd subbands.
(e) Schematic lattice structure of BPR with armchair edges. 
$N$ is the width of the ribbon. The translational vector $\bm{T}$ for armchair BPR is
 defined as $\bm{T} = T(0,1)$. 
(f) The 1st BZ of zigzag BPR (thick magenta line). The shaded yellow area is
 the 1st BZ of 2D BPN. 
In armchair BPR, its 1st BZ is projected
 onto $k_y$-axis and given as $-\pi/a_T\le k_y\le \pi/a_T$. 
The integration path $C$ for the calculation of the Zak phase is taken along
 $k_x$.
(g) Energy band structure for armchair BPR for $N=50$.
Cyan lines indicate the modes of TES that emerge in the energy gaps
 where the total Zak phase is $\pi$. 
(h) The corresponding Zak phase for armchair BPR for the lowest three subbands, 
where
Zak phase is finite in the $1$st and the $3$rd subbands.
  }
\label{fig:band_zak}
\end{figure*}
\begin{figure*}[t]
  \centering
  \includegraphics[width=0.9\linewidth]{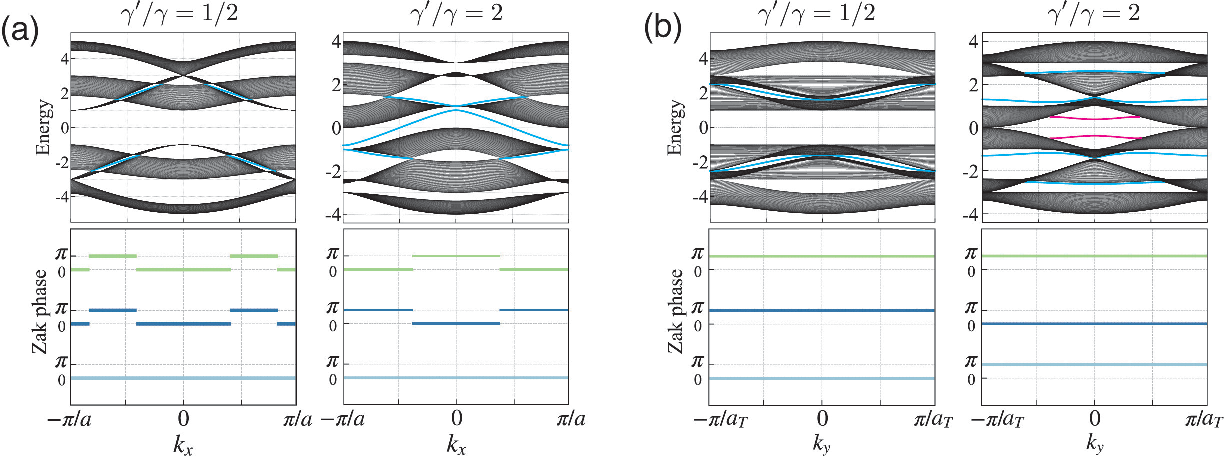}
  \caption{
     Energy band structure of BPR and the corresponding Zak phase. 
     (a) Zigzag-BPR with $\gamma^{\prime}/\gamma = 1/2$ (left) and $\gamma^{\prime}/\gamma = 2$ (right). 
In these case, TES (cyan lines) emerge in the energy gap where Zak phase has $\pi$. 
     (b) Armchair-BPR with $\gamma^{\prime}/\gamma = 1/2$  (left)  and
 $\gamma^{\prime}/\gamma = 2$ (right). 
TES with cyan lines appear in the energy gap that Zak phase has
 $\pi$. However, TES with magenta lines emerge in the energy band gap
 that Zak phase has $2\pi$. The puzzle will be resolved by considering
 Wannier centers (see main text). 
     }
\label{fig:zak_ribbon}
\end{figure*}
We numerically calculate Zak phase~\cite{Zak1989, King-Smith1993,Resta1994,Marzari2012,Zhou2015,Liu2017,Delplace2011} 
which is given as the line integral of Berry connection;
\begin{align}
{A}_{n} (k_{\parallel}, k_{\perp}) = i\bra{u_{n, \bm{k}}} \partial_{k_{\perp}} \ket{u_{n, \bm{k}}}.
\end{align}
The Berry connection physically means the vector potential in the reciprocal space.
In order to discuss the charge polarization based on the Zak phase, we have
introduced two specific one-dimensional (1D) wave numbers, i.e.,
$k_{\parallel}$ and $k_{\perp}$.
Here, $k_\parallel$ represents the 1D wave number 
parallel to the translational direction of the considered edge or ribbons. 
$k_{\perp}$ is a crystal momentum perpendicular to $k_\parallel$.

Zak phase of the $n$-th subband is defined as
\begin{equation}
  Z_n(k_\parallel) = \int_C dk_{\perp} {A}_{n}(k_{\parallel}, k_{\perp}),
\label{eq:zak}
\end{equation}
where $C$ is a straight path along $k_{\perp}$ connecting two equivalent points of $\bm{k}$ 
in the 2D BZ. For an inversion symmetric system, the Zak phase is quantized
by $0$ or $\pi$. 
Charge polarization for $n$-subband, $P_n$, is related with the Zak
phase as
\begin{equation}
P_n = \frac{1}{2\pi}Z_n.
\end{equation}
Thus, if the nontrivial Zak phase of $\pi$ appears, we obtain the finite charge
polarization at the surface or edge, which is nothing more than the
edge states.
Since the bulk topological properties are related to the charge
polarization at edges, this is referred to as bulk-edge correspondence. 
The Zak phase approach successfully predicts the topological
edge states (TES) for graphene, A3B nanosheet, 2D SSH model, and so
on~\cite{Ryu2002,Delplace2011,Kameda2019,Liu2017}.

Now, let us consider the 1D Biphenylene ribbons (BPRs). 
Figure~\ref{fig:band_zak} summarizes the lattice structure, electronic band structure and
Zak phase for BPRs.
To discuss the electronic properties of BPR, we shall define the
translationl vector of BPR as $\bm{T} = m\bm{a_1} + n\bm{a_2} = T(m, n)$,
and the corresponding reciprocal vector as 
 $\bm{\Gamma}_{\parallel} = 2\pi\frac{\bm{T}}{|\bm{T}|^2}$. 
Figure~\ref{fig:band_zak} (a) displays the schematic of BPR with zigzag
edges (zigzag-BPR),
where $N$ is the width of ribbons.
Since the translational vector of zigzag-BPR is given as 
$\bm{T} = T(1, 0)$, the 1st BZ of zigzag-BPR is given as $|k_x|\le
\pi/a$, which are marked as thick magenta lines in 
Fig.~\ref{fig:band_zak} (b).
The integration path $C$ for the calculation of the Zak phase is given in 
cyan arrow in Fig.~\ref{fig:band_zak} (b).
The energy band structure for zigzag-BPR with $N=50$ is shown in
Fig.~\ref{fig:band_zak} (c), where
black lines indicate the modes of bulk states, and cyan lines indicate the modes of
topological edge states, respectively.

For a numerical calculation, we rewrite Eq.~(\ref{eq:zak}) as a discrete form 
using Taylor expansion up to the first order, i.e.,
\begin{equation}
Z_n(k_{\parallel}) = -\rm{Im} \left[ \it{ln}
\left(\prod_{j=1}^{N_0}
\langle u_{n, \bm{k}_{\perp j}} | u_{n, \bm{k}_{\perp j+1}} \rangle
\right) \right],
\label{eq:zak_dis}
\end{equation} 
where $k_{\parallel}$ is a 1D momentum space parallel to $\bm{T}$. 
The integral path is divided into $N_{0}$ segments along $k_{\perp}$. 
Here, we impose for the wave function a gauge-fixing condition 
$| \braket{\phi_{g} | u_{n, k_{j}}} | \neq 0 $ in all $k_{j}$ point on the path
with a global gauge $\phi_{g}$, which is well-defined in the whole BZ~\cite{Fukui2005}.

The emergence of TES in zigzag-BPR can be associated with the nontrivial
Zak phase of $\pi$. Figure~\ref{fig:band_zak} (d) shows the $k$-dependent Zak phase of
zigzag-BPR for the lowest three bands. The Zak phase of the 1st subband is
always $0$ for any $k$, i.e., topological trivial. However, in the 2nd and
the 3rd subbands, there are $k$-regions ($|k|>k_c$) with finite Zak phase of $\pi$,
i.e., topologically nontrivial. Here,
$\pm k_c$ is the position of Dirac point between 2nd and 3rd subbands.
Thus, the TES emerge in the energy gap
between the 2nd and the 3rd subbands in the region of $|k|>k_c$, because $\sum_{n=1}^2 Z_n(k)=\pi$ for
$|k|<k_c$.
However, since $\sum_{n=1}^3 Z_n(k)=0$ for any $k$, there is no TES in
the energy gap between the 3rd and the 4th subbands. Other TES above $E>0$ are
also explained in a similar manner.

Next, we shall analyze the electronic and topological properties of BPR
with armchair edges (armchair-BPR). 
Figure~\ref{fig:band_zak} (e) displays the schematic of armchair-BPR, 
where $N$ is the width of the ribbon.
The translational vector of armchair-BPR is given as $\bm{T} = (0, 1)$.
Thus, the 1st BZ of armchair-BPR is given as $|k_y|\le
\pi/a_T$, which are marked as thick cyan line in 
Fig.~\ref{fig:band_zak} (f).
The energy band structure of armchair-BPR with $N=50$ is shown in
Fig.~\ref{fig:band_zak} (g), 
where the cyan lines in the energy gaps indicate TES modes.

The emergence of TES modes in armchair-BPR is also clearly explained by
the analysis of the Zak phase. 
Figure~\ref{fig:band_zak} (h) displays the $k$-dependent Zak phase for
the lowest three subbands of armchair-BPR, which gives $\pi$, $0$, $\pi$ from
the lowest subband.
Thus, the energy band gaps between the 1st and the 2nd subbands becomes
topologically nontrivial, because $Z_1(k) =
\pi$. Similarly, 
the energy band gaps between the 2nd and the 3rd subbands becomes
topologically nontrivial, because $\sum_{n=1}^2 Z_n(k) =
\pi$. These results are consistent with the emergence of TES shown in 
Figure~\ref{fig:band_zak} (g).
Other TES above $E>0$ are also explained similarly.

Similarly, from the viewpoint of Zak phase, 
it is illustrated that the existence of TES, even if we tune the ratio
of hopping energy, i.e., $\gamma^\prime/\gamma$.
Figures~\ref{fig:zak_ribbon} (a) and (b) display the energy band
 structures (upper) and the corresponding Zak phase (lower panel) for
 zigzag- and armchair-BPR, respectively, with the ratio 
 of hopping parameters tuned to $\gamma^\prime/\gamma=1/2$ and $2$.
For all cases of zigzag-BPR shown in Fig.~\ref{fig:zak_ribbon} (a),
the emergence of TES is consistent with the finite Zak phase of $\pi$.
For armchair-BPR shown in Fig.~\ref{fig:zak_ribbon} (b),
the emergence of TES for $\gamma^\prime/\gamma = 1/2$ (left panel) is
consistent with the presence of nontrivial Zak phase of $\pi$.
For $\gamma^\prime/\gamma = 2$ (right panel), 
the emergence of two TES modes with magenta lines in the gap between 
the 3rd and the 4th bands can not be illustrated solely by Zak phase, because
$\sum_{n=1}^3=2\pi$. Since the Zak phase of $2\pi$ is identical to the Zak
phase of $0$, TES should not appear. 
For armchair-BPR, in general, two TES modes appear in the band gap
between the 3rd and the 4th bands for $\gamma^\prime/\gamma > 1$, in spite of
that Zak phase is $2\pi$.
This puzzle can be explained by counting the Wannier centers using
elementary band representation in the following section.

\subsection{Elementary band representation}\label{ebr}
In the previous section, we showed that the Zak phase determines the existence of the edge states 
by numerical calculation. However, the ambiguity for the number of edge states has 
remained, because the Zak phase 
only gives the parities of $\mathbb{Z}_2$ value, e.g., the summation of
the Zak phase of the occupied bands becomes $0$ even if there are two
edge states in the gap. In order to obtain the accurate number of edge
states, we introduce the symmetry analysis using the irreducible
representation (irrep) at high-symmetric points of the 1st BZ
~\cite{Kruthoff2017,Bradlyn2017, Cano2018, Cano2020, Hitomi2021,
Tani2022}.  

The number of edge states is deeply involved with the central position of 
Wannier states (Wannier center) ~\cite{Marzari2012, Song2017, Kruthoff2017,Xu2021}. 
The edge terminated line crosses Wannier orbital $n$ times in a period
of the unit cell, 
$n$ edge states exist in the gap.  
The Wannier center is located at the midpoint of the bond with the stronger hopping energy. 
As discussed before, we considered two types of hopping energies between
two atomic sites, i.e., $\gamma$ and $\gamma^\prime$. 
Here, we identify the position of the Wannier center in terms of the
ratio $\gamma^\prime/\gamma$. 

To show the relationship between the position of the Wannier center and
the stronger hopping bond, we classify the energy bands of BPN by the
irreps of high-symmetric points in the 1st BZ. The space group of BPN is
$Pmmm$ which belongs to point group $D_{2h}$, and there are eight irreps
as shown in Table~\ref{tab:character_D2h}. Here, $C_{2j}$ ($j = x, y,
z$) is the 2-fold rotation around $j$-axis, $I$ is the inversion, 
$\sigma_{j}$ is the mirror reflection with respect to $kl$-plane ($j, k, l = x, y, z$). 
\begin{table}[ht]
 \caption{
  The character table of point group $D_{2h}$. 
  The four high-symmetric points in the 1st BZ belong to the pont goup $D_{2h}$. 
 }
 \label{tab:character_D2h}
 \centering
  \begin{tabular}{ccccccccc}
  \hline \hline
  $D_{2h}$ & $E$ & $C_{2z}$ & $C_{2y}$ & $C_{2x}$ & $I$ 
           & $\sigma_{z}$ & $\sigma_{y}$ & $\sigma_{x}$ \\
  \hline
  $\Gamma_{1}^{+}$ & 1 & 1 & 1 & 1 & 1 & 1 & 1 & 1 \\
  $\Gamma_{2}^{+}$ & 1 & 1 & -1 & -1 & 1 & 1 & -1 & -1 \\
  $\Gamma_{3}^{+}$ & 1 & -1 & 1 & -1 & 1 & -1 & 1 & -1 \\
  $\Gamma_{4}^{+}$ & 1 & -1 & -1 & 1 & 1 & -1 & -1 & 1 \\
  $\Gamma_{1}^{-}$ & 1 & 1 & 1 & 1 & -1 & -1 & -1 & -1 \\
  $\Gamma_{2}^{-}$ & 1 & 1 & -1 & -1 & -1 & -1 & 1 & 1 \\
  $\Gamma_{3}^{-}$ & 1 & -1 & 1 & -1 & -1 & 1 & -1 & 1 \\
  $\Gamma_{4}^{-}$ & 1 & -1 & -1 & 1 & -1 & 1 & 1 & -1 \\
  \hline \hline
  \end{tabular}
\end{table}

\begin{figure}[h]
 \centering
 \includegraphics[width=0.80\linewidth]{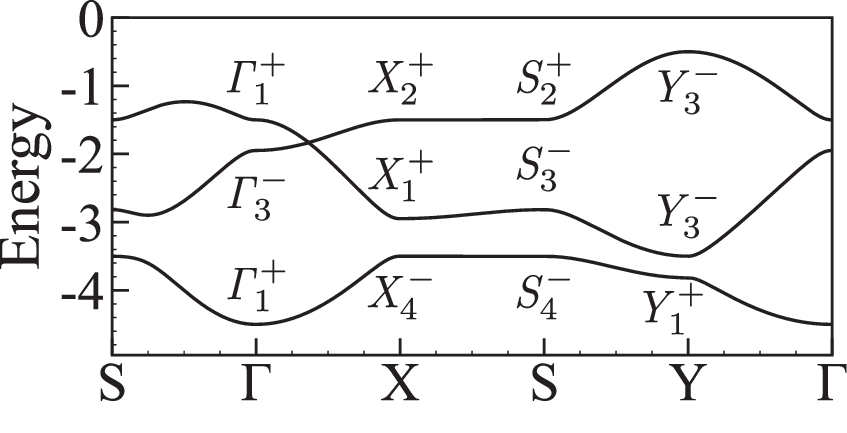}
 \caption{The irreps of the occupied energy bands for $\gamma^{\prime} >
 \gamma$. Note that only the lowest three energy bands of 2D BPN are displayed.}
 \label{fig:band_irrep}
\end{figure}
\begin{figure}[h]
  \centering
  \includegraphics[width=0.90\linewidth]{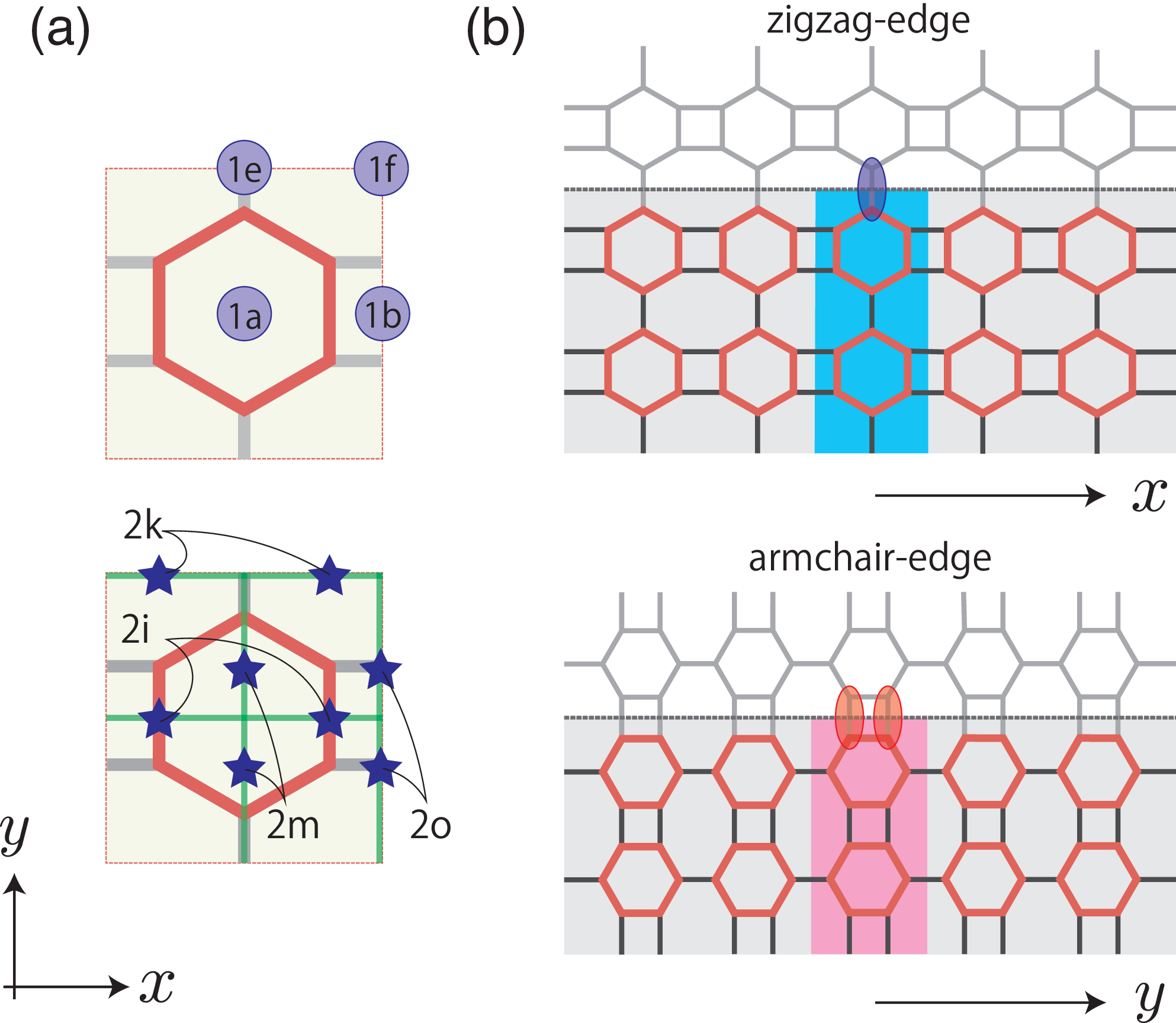}
  \caption{(a) Wyckoff positions (WPs) on $xy$-plane for the space group
 $Pmmm$. WPs $2i$, $2k$, $2m$, $2o$ are two positions on the green
 lines. In the case of $\gamma^{\prime}$ (gray line) > $\gamma$ (red
 line), the occupied three bands of BPN consist of two elementary bands
 $\bm{b}_{17}$ (WP $1e$) and $\bm{b}_{45}$ (WP $2o$) in
 Table~\ref{tab:ebr}. (b) (top) Since the edge terminated line of
 zigzag-BPR crosses Wannier orbital (indicated by the blue ellipse)
 once, one edge state appear. On the other hand, (bottom) the edge
 terminated line of armchair-BPR crosses Wannier orbitals (the red
 ellipse) twice, that result in two edge states. 
  }
  \label{fig:irrep_WP}
 \end{figure}
Each of the six bands in BPN is classified by the irreps at high-symmetric points of the 1st BZ. 
Since four high-symmetric points in the 1st BZ have the same symmetry as BPN lattice, 
the irreps at $\Gamma$, X, Y, and S points are identical to the point group $D_{2h}$. 
In Fig.~\ref{fig:band_irrep}, we show the correspondence between the occupied bands of BPN and 
the irreps characterized by the symmetry of the wave function.
From Fig.~\ref{fig:band_irrep}, we can define the character of the occupied bands by a single vector
\begin{equation}
  \begin{split}
  \bm{b} = (\gamma_{1}^{+}, \gamma_{2}^{+}, \gamma_{3}^{-}, \gamma_{4}^{-}; 
            x_{1}^{+}, x_{2}^{+}, x_{3}^{-}, x_{4}^{-}; \\ 
            y_{1}^{+}, y_{2}^{+}, y_{3}^{-}, y_{4}^{-}; 
            s_{1}^{+}, s_{2}^{+}, s_{3}^{-}, s_{4}^{-}), 
  \label{eq:ebr} 
  \end{split} 
\end{equation}
where $\gamma_{n}^{\pm}$, $x_{n}^{\pm}$, $y_{n}^{\pm}$, and $s_{n}^{\pm}$ 
with $n = 1, 2, \dots 4$ is the number of the bands 
whose wave function at $\Gamma$, X, Y, and S points have an irrep $\Gamma_{n}^{\pm}$. 
Here, we omitted the irreps 
$\Gamma_{3}^{+}$, $\Gamma_{4}^{+}$, $\Gamma_{1}^{-}$, and $\Gamma_{2}^{-}$
which are related to the parity with respect to the $xy$-plane, i.e., the
symmetry operation of $\sigma_z$,
because our tight-binding model is essentially assuming the $s$-orbital
as atomic orbitals which give always positive parity with respect to the
operation of $\sigma_z$ and become irrelevant. 
If we replace the atomic orbital on each site 
from $s$-orbital to $p_z$-orbital, we need to take into account
irreps $\Gamma_{3}^{+}$, $\Gamma_{4}^{+}$, $\Gamma_{1}^{-}$, and
$\Gamma_{2}^{-}$, but the discussion and the result will not be essentially changed.
Thus, the vector of the occupied $\pi$-bands of BPN is described as
\begin{equation}
  \bm{b}_{p_{z}} = (2, 0, 1, 0; 1, 1, 0, 1; 1, 0, 2, 0; 0, 1, 1, 1). 
\end{equation}

Here, we list all elementary bands allowed in the space group $Pmmm$ in Table \ref{tab:ebr}. 
An elementary band is a set of atomic orbitals situated at the 
symmetry protected
position, which is called a Wyckoff position (WP). 
The WP is classified by the site symmetry group (SSG), which guarantees
that the WP is invariant in the unit cell. 
In Fig.~\ref{fig:irrep_WP} (a), we show WPs in the unit cell. 
For instance, there are eight WPs which have the highest symmetry with the coordinates
$1a: (0, 0, 0)$, $1b: (a, 0, 0)$, $1c: (0, 0, z)$, $\dots$, $1h = (a, a_T, z)$, 
and their SSG is identical to the point group of the lattice, and has eight irreps. 
The WPs (and their coordinates) and the corresponding SSG are shown in
the first and the second column of Table ~\ref{tab:ebr}.  

Atomic orbitals located at WPs have irreps characterized by SSGs, which are the subgroups of 
the space group of the crystal. 
Each arrangement of the atomic orbitals is called the elementary band, 
and the irreps of the elementary band are elementary band representations (EBRs). 
In Table~\ref{tab:ebr}, we show all possible EBRs of WP on $xy$-plane. 
There are fifty different elementary bands with a serial number $\eta$ ($ = 1, 2, \dots , 50$). 

In the same manner as Eq.~(\ref{eq:ebr}), we shall describe the
elementary bands labeled by $\eta$ as 
\begin{equation}
  \begin{split}
    \bm{b}_{\eta} = (\gamma_{1}^{+(\eta)}, \gamma_{2}^{+(\eta)}, \gamma_{3}^{-(\eta)}, \gamma_{4}^{-(\eta)}; 
              x_{1}^{+(\eta)}, x_{2}^{+(\eta)}, x_{3}^{-(\eta)}, x_{4}^{-(\eta)}; \\ 
              y_{1}^{+(\eta)}, y_{2}^{+(\eta)}, y_{3}^{-(\eta)}, y_{4}^{-(\eta)}; 
              s_{1}^{+(\eta)}, s_{2}^{+(\eta)}, s_{3}^{-(\eta)}, s_{4}^{-(\eta)}). 
    \label{eq:ebr_eta} 
    \end{split} 
\end{equation}
Actual elements of $\bm{b}_{\eta}$ are listed from the 4th to the 19th
column in Table~\ref{tab:ebr},
where that Bloch function $e^{i{\bm{k}} \cdot {\bm{R}}}$ is taken into
account in each character. 
For example, the EBR of WP $1a$ with $\bm{R}=(0,0)$ gives identical
irreps among $\Gamma$, X, Y, S points.
In contrast, the EBR of WP 1b with $\bm{R}=(a,0)$ changes the sequence of
the irreducible representations at the X and S points depending on the
Bloch phase. 
The Wannier orbitals are the combination of the elementary bands, 
and their central positions are the corresponding WPs in the unit cell. 
As shown in the bottom row of Table ~\ref{tab:ebr}, 
the Wannier orbitals of occupied three bands of BPN can be described as
\begin{equation}
  \bm{b}_{p_{z}} = \sum_{\eta} n_{\eta} \bm{b}_{\eta}, 
\end{equation}
where $n_{\eta}$ is zero or positive number. 
From Table~\ref{tab:ebr}, $n_{\eta}$ is zero except for two EBRs;
$\eta = 17$ and $\eta = 45$ whose  WPs are$1e$ and $2o$, 
and correspond to SSGs ${mmm}$ and ${m2m}$ respectively. 
Therefore, the occupied bands of BPN are the linear combination of the
two elementary bands $\bm{b}_{17}$ and $\bm{b}_{45}$ 
which are both $s$-like orbitals situated at WPs $1e$ and $2o$. 
Hence, the Wannier orbitals are situated at the stronger hopping bonds, 
and the exact number of the edge states can be determined. 
Figure~\ref{fig:irrep_WP} (b) illustrates the broken Wannier orbitals 
at the zigzag and armchair edge boundaries for $\gamma^{\prime} > \gamma$. 
If the strength of the hopping energy is opposite ($\gamma^{\prime} < \gamma$), 
Wannier orbitals should be situated at the 6-membered rings (indicated
by the red line), not the edge boundaries. 
Recalling that $2\pi$ Zak phase leads to either
zero ($\gamma^{\prime} < \gamma$) or two ($\gamma^{\prime} > \gamma$) edge states in Sec.~\ref{sec:Zak}, 
it can be understood that the number of broken Wannier orbitals at the edge boundary 
consistent with the number of edge states.

\begin{table*}[ht]
  \caption{Elementary band representations (EBRs) of the space group $Pmmm$.} 
  \label{tab:ebr} 
  \centering
   \begin{tabular}{ccccccccccccccccccc|c}
   \hline \hline
   WP & SSG & irreps
   & $\gamma_{1}^{+}$ & $\gamma_{2}^{+}$ 
   & $\gamma_{3}^{-}$ & $\gamma_{4}^{-}$ 
   & $x_{1}^{+}$ & $x_{1}^{+}$ 
   & $x_{3}^{-}$ & $x_{4}^{-}$
   & $y_{1}^{+}$ & $y_{1}^{+}$ 
   & $y_{3}^{-}$ & $y_{4}^{-}$
   & $s_{1}^{+}$ & $s_{1}^{+}$ 
   & $s_{3}^{-}$ & $s_{4}^{-}$ & $\eta$ \\
   \hline
   \multirow{8}{*}{\shortstack{$1a$ \\ $(0, 0)$}} & \multirow{8}{*}{$mmm$} 
     & $\Gamma_{1}^{+}$ & 1 & 0 & 0 & 0 &   1 & 0 & 0 & 0 &   1 & 0 & 0 & 0 &   1 & 0 & 0 & 0 &  1\\
   & & $\Gamma_{2}^{+}$ & 0 & 1 & 0 & 0 &   0 & 1 & 0 & 0 &   0 & 1 & 0 & 0 &   0 & 1 & 0 & 0 &  2\\
   & & $\Gamma_{3}^{+}$ & 0 & 0 & 0 & 0 &   0 & 0 & 0 & 0 &   0 & 0 & 0 & 0 &   0 & 0 & 0 & 0 &  3\\
   & & $\Gamma_{4}^{+}$ & 0 & 0 & 0 & 0 &   0 & 0 & 0 & 0 &   0 & 0 & 0 & 0 &   0 & 0 & 0 & 0 &  4\\
   & & $\Gamma_{1}^{-}$ & 0 & 0 & 0 & 0 &   0 & 0 & 0 & 0 &   0 & 0 & 0 & 0 &   0 & 0 & 0 & 0 &  5\\
   & & $\Gamma_{2}^{-}$ & 0 & 0 & 0 & 0 &   0 & 0 & 0 & 0 &   0 & 0 & 0 & 0 &   0 & 0 & 0 & 0 &  6\\
   & & $\Gamma_{3}^{-}$ & 0 & 0 & 1 & 0 &   0 & 0 & 1 & 0 &   0 & 0 & 1 & 0 &   0 & 0 & 1 & 0 &  7\\
   & & $\Gamma_{4}^{-}$ & 0 & 0 & 0 & 1 &   0 & 0 & 0 & 1 &   0 & 0 & 0 & 1 &   0 & 0 & 0 & 1 &  8\\ 
   \hline
   \multirow{8}{*}{\shortstack{$1b$ \\ $(a, 0)$}} & \multirow{8}{*}{$mmm$} 
     & $\Gamma_{1}^{+}$  & 1 & 0 & 0 & 0 &   0 & 0 & 0 & 1 &   1 & 0 & 0 & 0 &   0 & 0 & 0 & 1 &  9\\
   & & $\Gamma_{2}^{+}$ & 0 & 1 & 0 & 0 &   0 & 0 & 1 & 0 &   0 & 1 & 0 & 0 &   0 & 0 & 1 & 0 &  10\\
   & & $\Gamma_{3}^{+}$ & 0 & 0 & 0 & 0 &   0 & 0 & 0 & 0 &   0 & 0 & 0 & 0 &   0 & 0 & 0 & 0 &  11\\
   & & $\Gamma_{4}^{+}$ & 0 & 0 & 0 & 0 &   0 & 0 & 0 & 0 &   0 & 0 & 0 & 0 &   0 & 0 & 0 & 0 &  12\\
   & & $\Gamma_{1}^{-}$  & 0 & 0 & 0 & 0 &   0 & 0 & 0 & 0 &   0 & 0 & 0 & 0 &   0 & 0 & 0 & 0 &  13\\
   & & $\Gamma_{2}^{-}$ & 0 & 0 & 0 & 0 &   0 & 0 & 0 & 0 &   0 & 0 & 0 & 0 &   0 & 0 & 0 & 0 &  14\\
   & & $\Gamma_{3}^{-}$ & 0 & 0 & 1 & 0 &   0 & 1 & 0 & 0 &   0 & 0 & 1 & 0 &   0 & 1 & 0 & 0 &  15\\
   & & $\Gamma_{4}^{-}$ & 0 & 0 & 0 & 1 &   1 & 0 & 0 & 0 &   0 & 0 & 0 & 1 &   1 & 0 & 0 & 0 &  16\\ 
   \hline
   \multirow{8}{*}{\shortstack{$1e$ \\ $(0, a_{T})$}} & \multirow{8}{*}{$mmm$} 
     & $\Gamma_{1}^{+}$  & 1 & 0 & 0 & 0 &   1 & 0 & 0 & 0 &   0 & 0 & 1 & 0 &   0 & 0 & 1 & 0 &  17\\
   & & $\Gamma_{2}^{+}$ & 0 & 1 & 0 & 0 &   0 & 1 & 0 & 0 &   0 & 0 & 0 & 1 &   0 & 0 & 0 & 1 &  18\\
   & & $\Gamma_{3}^{+}$ & 0 & 0 & 0 & 0 &   0 & 0 & 0 & 0 &   0 & 0 & 0 & 0 &   0 & 0 & 0 & 0 &  19\\
   & & $\Gamma_{4}^{+}$ & 0 & 0 & 0 & 0 &   0 & 0 & 0 & 0 &   0 & 0 & 0 & 0 &   0 & 0 & 0 & 0 &  20\\
   & & $\Gamma_{1}^{-}$  & 0 & 0 & 0 & 0 &   0 & 0 & 0 & 0 &   0 & 0 & 0 & 0 &   0 & 0 & 0 & 0 &  21\\
   & & $\Gamma_{2}^{-}$ & 0 & 0 & 0 & 0 &   0 & 0 & 0 & 0 &   0 & 0 & 0 & 0 &   0 & 0 & 0 & 0 &  22\\
   & & $\Gamma_{3}^{-}$ & 0 & 0 & 1 & 0 &   0 & 0 & 1 & 0 &   1 & 0 & 0 & 0 &   1 & 0 & 0 & 0 &  23\\
   & & $\Gamma_{4}^{-}$ & 0 & 0 & 0 & 1 &   0 & 0 & 0 & 1 &   0 & 1 & 0 & 0 &   0 & 1 & 0 & 0 &  24\\ 
   \hline
   \multirow{8}{*}{\shortstack{$1f$ \\ $(a, a_{T})$}} & \multirow{8}{*}{$mmm$} 
     & $\Gamma_{1}^{+}$  & 1 & 0 & 0 & 0 &   0 & 0 & 0 & 1 &   0 & 0 & 1 & 0 &   1 & 0 & 0 & 0 &  25\\
   & & $\Gamma_{2}^{+}$ & 0 & 1 & 0 & 0 &   0 & 0 & 1 & 0 &   0 & 0 & 0 & 1 &   0 & 1 & 0 & 0 &  26\\
   & & $\Gamma_{3}^{+}$ & 0 & 0 & 0 & 0 &   0 & 0 & 0 & 0 &   0 & 0 & 0 & 0 &   0 & 0 & 0 & 0 &  27\\
   & & $\Gamma_{4}^{+}$ & 0 & 0 & 0 & 0 &   0 & 0 & 0 & 0 &   0 & 0 & 0 & 0 &   0 & 0 & 0 & 0 &  28\\
   & & $\Gamma_{1}^{-}$  & 0 & 0 & 0 & 0 &   0 & 0 & 0 & 0 &   0 & 0 & 0 & 0 &   0 & 0 & 0 & 0 &  29\\
   & & $\Gamma_{2}^{-}$ & 0 & 0 & 0 & 0 &   0 & 0 & 0 & 0 &   0 & 0 & 0 & 0 &   0 & 0 & 0 & 0 &  30\\
   & & $\Gamma_{3}^{-}$ & 0 & 0 & 1 & 0 &   0 & 1 & 0 & 0 &   1 & 0 & 0 & 0 &   0 & 0 & 1 & 0 &  31\\
   & & $\Gamma_{4}^{-}$ & 0 & 0 & 0 & 1 &   1 & 0 & 0 & 0 &   0 & 1 & 0 & 0 &   0 & 0 & 0 & 1 &  32\\ 
   \hline
   \multirow{4}{*}{\shortstack{$2i$ \\ $(\pm{x}, 0)$}} & \multirow{4}{*}{$2mm$} 
     & $\Gamma_{1}$ & 1 & 0 & 0 & 1 &   1 & 0 & 0 & 1 &   1 & 0 & 0 & 1 &   1 & 0 & 0 & 1 &  33\\
   & & $\Gamma_{2}$ & 0 & 0 & 0 & 0 &   0 & 0 & 0 & 0 &   0 & 0 & 0 & 0 &   0 & 0 & 0 & 0 &  34\\
   & & $\Gamma_{3}$ & 0 & 1 & 1 & 0 &   0 & 1 & 1 & 0 &   0 & 1 & 1 & 0 &   0 & 1 & 1 & 0 &  35\\
   & & $\Gamma_{4}$ & 0 & 0 & 0 & 0 &   0 & 0 & 0 & 0 &   0 & 0 & 0 & 0 &   0 & 0 & 0 & 0 &  36\\
   \hline
   \multirow{4}{*}{\shortstack{$2k$ \\ $(\pm{x}, a_{T})$}} & \multirow{4}{*}{$2mm$} 
     & $\Gamma_{1}$& 1 & 0 & 0 & 1 &   1 & 0 & 0 & 1 &   0 & 1 & 1 & 0 &   0 & 1 & 1 & 0 &  37\\
   & & $\Gamma_{2}$ & 0 & 0 & 0 & 0 &   0 & 0 & 0 & 0 &   0 & 0 & 0 & 0 &   0 & 0 & 0 & 0 &  38\\
   & & $\Gamma_{3}$ & 0 & 1 & 1 & 0 &   0 & 1 & 1 & 0 &   1 & 0 & 0 & 1 &   1 & 0 & 0 & 1 &  39\\
   & & $\Gamma_{4}$ & 0 & 0 & 0 & 0 &   0 & 0 & 0 & 0 &   0 & 0 & 0 & 0 &   0 & 0 & 0 & 0 &  40\\
   \hline
   \multirow{4}{*}{\shortstack{$2m$ \\ $(0, \pm{y})$}} & \multirow{4}{*}{$m2m$} 
     & $\Gamma_{1}$ & 1 & 0 & 1 & 0 &   1 & 0 & 1 & 0 &   1 & 0 & 1 & 0 &   1 & 0 & 1 & 0 &  41\\
   & & $\Gamma_{2}$ & 0 & 0 & 0 & 0 &   0 & 0 & 0 & 0 &   0 & 0 & 0 & 0 &   0 & 0 & 0 & 0 &  42\\
   & & $\Gamma_{3}$ & 0 & 1 & 0 & 1 &   0 & 1 & 0 & 1 &   0 & 1 & 0 & 1 &   0 & 1 & 0 & 1 &  43\\
   & & $\Gamma_{4}$ & 0 & 0 & 0 & 0 &   0 & 0 & 0 & 0 &   0 & 0 & 0 & 0 &   0 & 0 & 0 & 0 &  44\\
   \hline
   \multirow{4}{*}{\shortstack{$2o$ \\ $(a, \pm{y})$}} & \multirow{4}{*}{$m2m$} 
     & $\Gamma_{1}$ & 1 & 0 & 1 & 0 &   0 & 1 & 0 & 1 &   1 & 0 & 1 & 0 &   0 & 1 & 0 & 1 &  45\\
   & & $\Gamma_{2}$ & 0 & 0 & 0 & 0 &   0 & 0 & 0 & 0 &   0 & 0 & 0 & 0 &   0 & 0 & 0 & 0 &  46\\
   & & $\Gamma_{3}$ & 0 & 1 & 0 & 1 &   1 & 0 & 1 & 0 &   0 & 1 & 0 & 1 &   1 & 0 & 1 & 0 &  47\\
   & & $\Gamma_{4}$ & 0 & 0 & 0 & 0 &   0 & 0 & 0 & 0 &   0 & 0 & 0 & 0 &   0 & 0 & 0 & 0 &  48\\
   \hline
   \multirow{2}{*}{\shortstack{$4y$ \\ $(\pm{x}, \pm{y})$}} & \multirow{2}{*}{$..m$} 
     & $\Gamma_{1}$          & 1 & 1 & 1 & 1 &   1 & 1 & 1 & 1 &  1 & 1 & 1 & 1 &   1 & 1 & 1 & 1 &  49\\
   & & $\Gamma_{2}$ & 0 & 0 & 0 & 0 &   0 & 0 & 0 & 0 &  0 & 0 & 0 & 0 &   0 & 0 & 0 & 0 &  50\\
   \hline
   \multicolumn{3}{c}{$\bm{b}_{p_{z}}$ (occupied $p_{z}$ bands)}
    & 2 & 0 & 1 & 0 &    1 & 1 & 0 & 1 &    1 & 0 & 2 & 0 &    0 & 1 & 1 & 1 &  17 + 45\\
   \hline \hline
 \end{tabular}
 \end{table*}

\subsection{Topological Corner States}\label{sec:corner} 
\begin{figure}[t]
 \centering
 \includegraphics[width=0.9\linewidth]{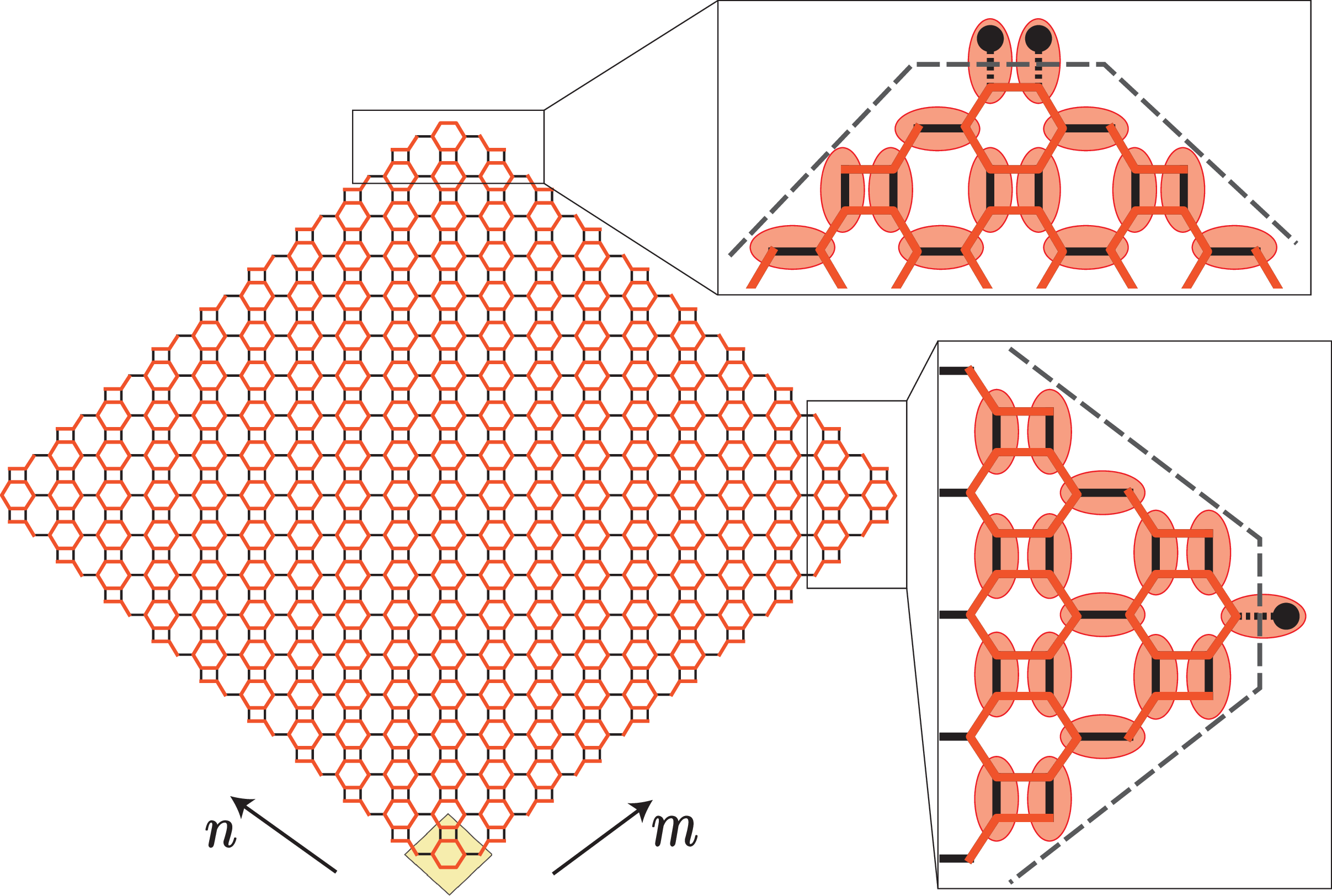}
 \caption{
  The structure of the BPN nanoflake for $\gamma^{\prime} > \gamma$. 
  The red ellipses indicate the Wannier orbitals. 
  Deleting the atom basis indicated black dots at the four corners, 
  there are ($m \times n \times 12 - 6$) atom sites. 
  Wannier orbitals are cut by the corner geometries 
  resulting in uncoupled Wannier orbitals at the corner sites. 
 }
 \label{fig:corner_structure}
\end{figure}
\begin{figure*}[t]
  \centering
  \includegraphics[width=0.95\textwidth]{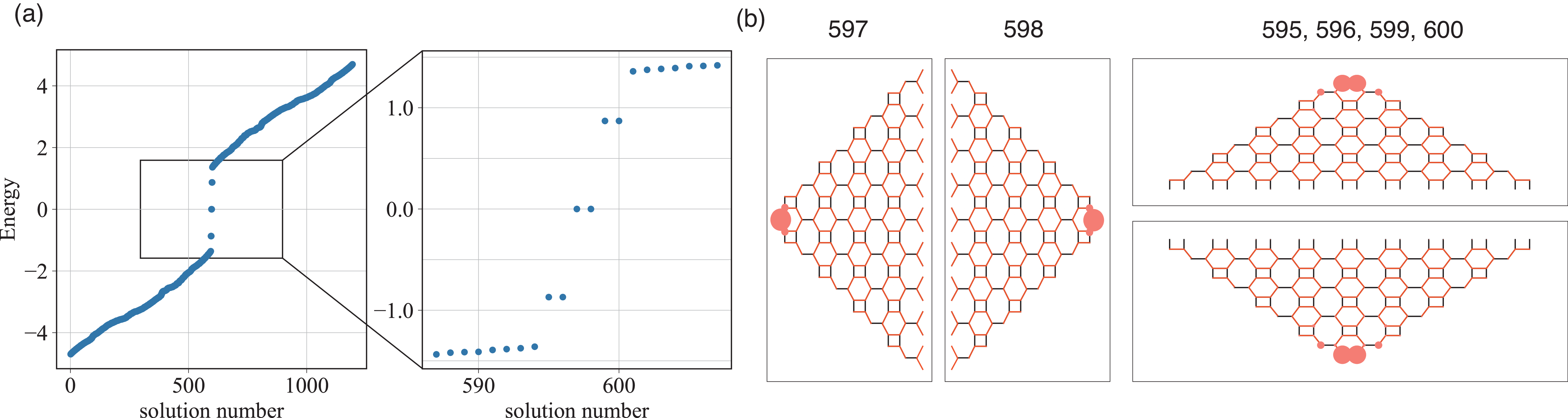}
  \caption{
    (a) The energy spectrum of the BPN nanoflake. 
    (b) The amplitude of the wave function of the six energy levels in the gap. 
        Two states are localized at the left and right corner, 
        and the other four states are localized at the upper and lower corner. }
  \label{fig:corner_ene_wf}
 \end{figure*}
In this section, we consider BPN nanoflake to investigate the conditions 
for the appearance of topological corner states as "edge-of-edge"
states~\cite{Langbehn2017, Hashimoto2017, Ezawa2018, Wladimir2017,
Liu2021}.  
In a similar manner to that shown in the previous section, 
we assume a geometry in which 
the Wannier orbitals are isolated at the boundary of the system. 
Figure~\ref{fig:corner_structure} illustrates 
the structure of the BPN nanoflake formed as a combination of diagonally cut lines. 
The rhombus shaded by yellow is a unit, which is 
periodic in the $m$ and $n$ directions.
Here, the red lines indicate the hopping of $\gamma$ 
and the other black lines indicate the hopping $\gamma^{\prime}$. 
For $\gamma^{\prime} > \gamma$, edge states exist in the energy gap,
where the Wannier orbitals are located on the bonds indicated by the red
ellipses as shown 
in the zoom of Figure~\ref{fig:corner_structure}.
Furthermore, if the basis of atomic sites marked by the black circles at
the four corners is removed, the Wannier orbitals are broken at the
corner sites, i.e., the emergence of corner states.

Figure~\ref{fig:corner_ene_wf} (a) shows the energy spectrum of the nanoflake,
and the right figure is a zoomed-in view near zero energy. 
Since the system includes 1194 atoms in the case of $(m, n) = (10, 10)$, 
the site index has that same number. 
For $\gamma^{\prime} > \gamma$, it has an energy gap near zero energy, 
and six isolated energy levels are found within the gap. 
Figure~\ref{fig:corner_ene_wf} (b) is the wave function of the six states 
within the gap near zero energy. 
We plot the amplitude of the wave function as the radius of the circles. 
We can see the exponentially localized states at the corners of the nanoflake. 
Two states with $E = 0$ are localized in the left and right corners, 
while the other four states are localized in the lower and upper corners. 
This can be understood from the position of the Wannier orbitals. 
The Wannier orbitals cut by the corner geometry are present: one at the
left and right corners and two at the upper and lower corners, 
resulting in a corresponding number of corner states. 
Therefore, we can see that these corner states are due to 
the uncoupled Wannier orbitals at the corners of the system. 
In addition, changing the values of $m$ and $n$ does not affect 
the appearance of corner states.

\section{conclusion}\label{sec:conclusion}
In this paper, we have studied the electronic and topological
properties of BPN. We have demonstrated that BPN exhibits flat bands and
tilted Dirac cones, and these tilted cones persist even when the
strength of electron hoppings is modified. By varying the ratio of
hopping energies, specifically the ratio between intra-cellular hopping
$\gamma$ and inter-cellular hopping $\gamma^{\prime}$, we can open an
energy band gap, leading to a topological phase transition. The
emergence of edge states is guaranteed by a nonzero Zak phase, and the
number of edge states aligns with the number of Wannier orbitals at the
edge boundary. Using the same principle, we have illustrated the
existence of corner states in systems where uncoupled Wannier orbitals
are located at the corners. The present theoretical approach is
applicable to the design of new materials that share the same structure
as BPNs. Although it is very hard to tune the electron hopping energies in actual
BPN materials, we can design the system having the topological
properties of BPN based on photonic crystals. The extension to
the photonic crystals will be discussed elsewhere.

\begin{acknowledgments}
The authors are grateful to M. Hitomi for helpful discussions.
This work was supported by JSPS KAKENHI
(Nos. 22H05473, JP21H01019, JP18H01154) and JST CREST (No. JPMJCR19T1).
\end{acknowledgments}

\appendix
\section{tight-binding parameters of 2D BPN}
\begin{figure*}[t]
  \centering
  \includegraphics[width=0.95\textwidth]{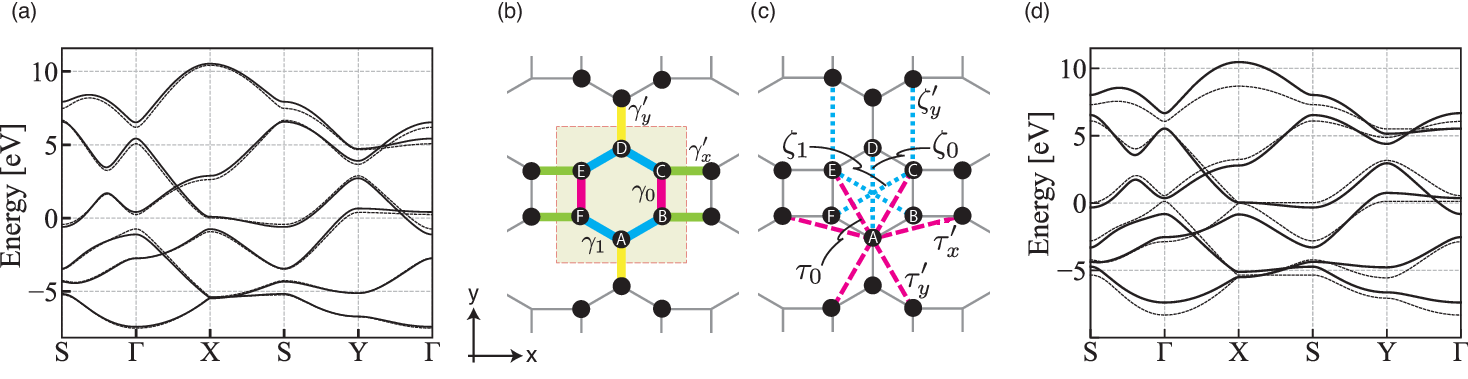}
  \caption{
(a) Energy band structure obtained by DFT (dashed line) and Wannier90
 (solid line). In Wanner90 calculation, only p$_z$ orbitals of carbon
 atoms are taken into account for the projection. 
(b) Schematic of BPN lattice structure with
the definition of nearest neighbor (n.n.) hoppings
for the tight-binding
 model of 2D BPN. 
Yellow shaded rectangle is unit cell. 
$\gamma_0$ is 
 intra-cellular n.n. hoppings for the bond of B-C and E-F. 
$\gamma_0$ is intra-cellular n.n. hoppings for the bond of A-B, C-D, D-E
 and F-A. 
$\gamma^\prime_x$ and
$\gamma^\prime_y$ are the inter-cellular hoppings for $x$- and
 $y$-directions,
respectively. 
(c) The definition of 2nd n.n. and 3rd n.n. electron hoppings.
Magenta thick dashed lines indicate the 2nd n.n. hoppings. $\tau_0$
 is intra-cellular hoppings for the bond of A-E, A-C. $\tau^\prime_y$ is
 inter-cellular hoppings for the bond of A-E, A-C. 
$\tau^\prime_x$ is inter-cellular hoppings for the bond of A-B, A-F. 
Cyan dot lines indicate the 3rd n.n. hoppings.
$\zeta_0$ is intra-cellular hoppings for the bond of A-D. 
$\zeta_1$ is intra-cellular hoppings for the bond of B-E and C-F. 
$\zeta^\prime_y$ is inter-cellular hoppings for the bond of B-C and E-F. 
Actual values of the hoppings are summarized in Table~\ref{tab:hopping}. 
(d) Energy band structure of effective tight-binding model considering
    up to the n.n. hopping (dashed line) and the 3rd n.n. hopping (solid
 line).  
Inclusion up to 3rd n.n. hoppings sufficiently well reproduces the energy band
 structures obtained by DFT shown in (a).
}
  \label{fig:w90}
 \end{figure*}

In this section, we present the parameter set for the tight-binding model
to describe $\pi$-electronic states of 2D BPN. 
We have implemented the first-principles calculations to obtain the
energy band structure for 2D BPN based on the density functional theory (DFT). 
To this end, we have used the QunatumEspresso package~\cite{qe2009,qe2017}
by employing the PBE variant of the GGA method~\cite{PBE}. 
A relaxation of the structure was achieved by considering a magnitude
of the forces on each atom that is less than $13.6\times 10^{-4}$ eV/\AA. 
A $12\times 14 \times 1$ Monkhorst–Pack mesh~\cite{MKmethod} is used to sample the 1st
BZ for structural relaxation. 
Figure~\ref{fig:w90}(a) shows the obtained $\pi$-electronic energy band
structures of 2D BPN (dashed line).

In further, we have extracted the tight-binding parameters for
$\pi$-electrons of 2D BPN using the Wannier90 package~\cite{wannier90}.  
The parameter set reproduces quite well the energy band structure
obtained by DFT as shown in Fig.~\ref{fig:w90} (a), where the solid lines
are the tight-binding calculation obtained by Wannier90. 

Figures~\ref{fig:w90} (b) and (c) are schematic for the positions of
the hopping integrals for tight-binding model, 
where the nearest neighbor (n.n.) hoppings are shown in
Fig~\ref{fig:w90}(b).
Similarly, the 2nd and the 3rd n.n. hoppings are shown in Fig~\ref{fig:w90}(c). 
The obtained electron hopping parameters are listed in Table~\ref{tab:hopping}.
Figure~\ref{fig:w90}(d) shows the energy band structure of
$\pi$-electrons for 2D BPN. The dashed lines indicate the band
structure, which only includes n.n. hoppings. However, the solid lines
indicate the energy band structure, which includes up to 3rd
n.n. hoppings. The inclusion up to 3rd n.n. hoppings sufficiently well
reproduce the energy band structure obtained by DFT.
Especially, 3rd n.n. hoppings are necessary to reproduce 
the band crossing between 3rd and 4th subbands
on the Y-$\Gamma$ line, the band gap opening at X and the slight tilt of
the flat band on X-S line. 
\begin{table*}[t]
  \caption{
Tight-binding parameters for 2D BPN. The values are derived from the
 Wannier90 calculation. The set of these parameters well reproduces the
 energy band structures of $\pi$-electrons for 2D BPN as shown in
 Fig.~\ref{fig:w90} (d). 
   The list of the hopping integrals. 
$V_{\alpha}$ ($V_{\beta}$) is the
 on-site potential energy on A, D (B, C, E, F) sites, respectively. 
The definition electron hoppings on the lattice of 2D BPN are shown in
 Figs.~\ref{fig:w90} (b) and (c). 
  }
  \label{tab:hopping}
  \centering
   \begin{tabular}{cc|cccc|ccc|ccc}
   \hline\hline
   \multicolumn{2}{c}{potential} & \multicolumn{4}{c}{n.n.} & \multicolumn{3}{c}{2nd n.n.} &\multicolumn{3}{c}{3rd n.n.}\\
   \hline
   $V_{\alpha}$ & $V_{\beta}$ & $\gamma_{0}$ & $\gamma_{1}$ & $\gamma^{\prime}_{x}$ & $\gamma^{\prime}_{y}$ & $\tau_{0}$ & $\tau^{\prime}_{x}$ & $\tau^{\prime}_{y}$ & $\zeta_{0}$ & $\zeta{1}$ & $\zeta^{\prime}_{y}$ \\
   -2.0742 & -2.3647 
   & -2.9603 & -2.6910 & -2.7479 & -2.7962 
   & 0.2880 & 0.3969 & 0.2473 
   & -0.2041 & -0.2991 & -0.1920 \\
   \hline \hline
   \end{tabular}
 \end{table*}

\bibliographystyle{apsrev4-1}
\bibliography{reference}

\end{document}